\documentclass[
    ,final            
  ]
  {aipproc}

\layoutstyle{6x9}

\usepackage{graphicx}
\usepackage{epsf}

\usepackage{amsmath, amssymb}

\begin{document}

\begin{center}
\hfill TU-859 \\
\hfill UT-HET 032 \\
\hfill December, 2009
\end{center}
\title{Cosmic Gamma-ray from Inverse Compton Process in
Unstable Dark Matter Scenario}


\author{Koji Ishiwata}{
  address={Department of Physics, Tohoku University,
    Sendai 980-8578, Japan}
}

\author{ Shigeki Matsumoto}{
  address={Department of Physics, University of Toyama, 
    Toyama 930-8555, Japan}
}

\author{Takeo Moroi}{
  address={Department of Physics, Tohoku University,
    Sendai 980-8578, Japan}
}

\begin{abstract}
  Motivated by the PAMELA anomaly in the fluxes of cosmic-ray $e^+$ and
  $e^-$, we study the cosmic $\gamma$-ray induced by the inverse Compton
  (IC) scattering process in unstable dark matter scenario assuming that
  the anomaly is due to the $e^\pm$ emission by the decay of dark
  matter.  We calculate the fluxes of IC-induced $\gamma$-ray produced
  in our Galaxy and that from cosmological distance, and show that both
  of them are significant.  We discuss a possibility that large dark
  matter mass over TeV scale might be constrained by the $\gamma$-ray
  observation by Fermi Gamma-ray Space Telescope.

\end{abstract}

\maketitle



\section{Introduction}


Recent observations of the fluxes of high-energy cosmic rays have made
an impact on the understanding of the nature of dark matter (DM).  In
particular, the PAMELA experiment has observed an increasing behavior of
the positron fraction in the cosmic ray in the energy range of $10\ {\rm
GeV}\lesssim E_e\lesssim 100\ {\rm GeV}$ (with $E_e$ being the energy of
$e^\pm$) \cite{Adriani:2008zr}, which cannot be explained if we consider
the conventional $e^\pm$ fluxes in astrophysics.  This fact suggests
that there may exist a non-standard source of energetic positron (and
electron) in our Galaxy.

One of the possibilities is unstable dark matter.  If dark matter has
lifetime of $O(10^{25}-10^{26}\ {\rm sec})$, and also if positron is
produced by the decay, the PAMELA anomaly may be explained; for early
attempts, see, for example, \cite{Huh:2008vj, Nomura:2008ru,
  Yin:2008bs, Ishiwata:2008cv, Bai:2008jt, Chen:2008md,
  Hamaguchi:2008rv, Ponton:2008zv, Ibarra:2008jk, Chen:2008qs,
  Arvanitaki:2008hq, Hamaguchi:2008ta, Chen:2008dh}.  (Other
possibilities of explaining the PAMELA anomaly include the dark-matter
annihilation \cite{Bergstrom:2008gr, Cirelli:2008pk, Cholis:2008qq,
  Feldman:2008xs, Fox:2008kb, Barger:2008su, Nelson:2008hj,
  Harnik:2008uu, Ibe:2008ye, Hooper:2008kv} and the positron emission
from pulsars \cite{Hooper:2008kg}.)  In addition, a precise
measurement of the total $(e^++e^-)$ flux has been performed by the
Fermi Gamma-ray Space Telescope \cite{Abdo:2009zk}, whose results
suggest that the flux of $(e^++e^-)$ is proportional to $\sim
E_e^{-3}$.  Although the interpretation of the Fermi result is still
controversial, it has been discussed that the observed spectrum may be
too hard to be consistent with the prediction of conventional
astrophysical model, and that the $e^\pm$ observed by the Fermi
experiment may be significantly contaminated by the $e^\pm$ produced
by the decay of dark matter.  Such a scenario suggests the mass of
$m_{\rm DM}\sim O(1\ {\rm TeV})$ and the lifetime of $\tau_{\rm
  DM}\sim O(10^{26}\ {\rm sec})$ \cite{Meade:2009iu, Shirai:2009fq,
  MarNomYha}.

If the decay of dark matter is the source of the extra positrons
observed by the PAMELA experiment, the emitted positron and electron
produce photon via synchrotron radiation and inverse Compton (IC)
scattering.  In our Galaxy, energy loss rates of the energetic $e^\pm$
via these processes are of the same order, but the typical energy of
the photon emitted by these processes is different.  Since the
magnetic field in our Galaxy is expected to be $O(1\ \mu{\rm G})$, the
energy of the synchrotron radiation is typically $10^{-3}\ {\rm eV}$
when the energy of $e^\pm$ is $O(1\ {\rm TeV})$.  (The synchrotron
radiation from the Galactic center is discussed in \cite{Nardi:2008ix,
  Ishiwata:2008qy}.)  On the contrary, the IC process produces
$\gamma$-rays with higher energy.  If an energetic $e^\pm$ with
$E_e\sim O(1\ {\rm TeV})$ scatters off the cosmic microwave background
(CMB) photon, $\gamma$-ray with $E_\gamma\sim O(1-10\ {\rm GeV})$ is
produced.  In addition, in our Galaxy, there exist background photons
from stars, which have higher energy than the CMB radiation.  The IC
scattering with those photons produces $\gamma$-ray with higher
energy.  Importantly, the high-energy $\gamma$-ray flux can be
precisely measured by the Fermi telescope.  Thus, in order to examine
the scenario in which the PAMELA anomaly is explained by the decay of
dark matter, it is important to study the energetic $\gamma$-ray
emitted by the IC process.

In this paper, we study the flux of $\gamma$-ray produced by the IC
process in the decaying dark matter scenario.  We pay particular
attention to the parameter space in which the positron fraction is in
agreement with the PAMELA results.  In our previous work
\cite{Ishiwata:2009dk}, IC-induced $\gamma$-ray in various direction
from Galactic and extra-Galactic region are shown; we have seen that the
extra-Galactic contribution can be as large as Galactic one.  In
particular, if we see $\gamma$-ray in the direction off the Galactic
center, signal of the IC-induced $\gamma$-ray from extra-Galactic region
may be observed.  Here, paying attention to the observation data by
Fermi, we compare theoretically-calculated $\gamma$-ray with the
observations.
It will be shown that the parameter region $m_{\rm DM} \gtrsim 4$ TeV,
which is suggested by PAMELA and Fermi observations, might be
constrained by the observation of isotropic $\gamma$-ray by Fermi
\cite{Abdo:2009ka}. We will also show that the flux of the IC-induced
$\gamma$-ray in our Galaxy may be less than the observed flux in
intermediate Galactic latitude (IGL) given by Fermi \cite{Porter:2009sg}
when $\tau_{\rm DM}\sim O(10^{26}\ {\rm sec})$ to explain the PAMELA
anomaly.

\section{Formula of $\gamma$-ray in inverse-Compton process}

We first discuss the procedure to calculate the $\gamma$-ray flux. The
following formula is based on \cite{Ishiwata:2009dk}.  

The total $\gamma$-ray flux is given by the sum of two contributions:
\begin{eqnarray}
  \Phi_{\gamma_{\rm IC}} = 
  \Phi_{\gamma}^{\rm (Galaxy)} + 
  \Phi_{\gamma}^{\rm (Cosmo)}, 
\end{eqnarray}
where the first and second terms in the right-hand side are fluxes of
$\gamma$-ray produced in our Galaxy and that from cosmological
distance (i.e., extra-Galactic contribution), respectively.  Notice
that $\Phi_{\gamma}^{\rm (Cosmo)}$ is isotropic, while
$\Phi_{\gamma}^{\rm (Galaxy)}$ depends on direction we observe.

In order to discuss the IC-induced $\gamma$-ray, it is necessary to
understand the spectrum of the parent $e^\pm$.  In our Galaxy, energetic
$e^\pm$ is approximately in a random-walk motion because of the
entangled magnetic field.  Then, the $e^{\pm}$ energy spectrum $f_e$
(i.e., number density of ($e^++e^-$) per unit energy) in our Galaxy is
described by the following diffusion equation:
\begin{eqnarray}
  K_e(E_e) \nabla^2 f_{e}(E,\vec{x})
  + \frac{\partial}{\partial E}
  \left[ b_{\rm loss}(E_e,\vec{x}) f_{e}(E_e,\vec{x}) \right]
  + Q_e(E_e,\vec{x}) = 0,
  \label{DiffEq}
\end{eqnarray}
where $K_E(E_E)$ is the diffusion coefficient, $b_{\rm
loss}(E_e,\vec{x})$ is the energy loss rate, and $Q_e(E_e,\vec{x})$ is
the $e^{\pm}$ source term.  In considering the long-lived dark matter,
the source term is given by
\begin{eqnarray}
  Q_e (E_e,\vec{x})= \frac{1}{\tau_{\rm DM}}
  \frac{\rho^{\rm (Galaxy)}_{\rm DM}(\vec{x})}{m_{\rm DM}}
  \frac{dN_{e}}{dE_e},
  \label{sourceterm}
\end{eqnarray}
where $\rho_{\rm DM}^{\rm (Galaxy)}$ is energy density of dark matter
and $dN_{e}/dE_e$ is energy distribution of $e^\pm$ from the decay of
single dark matter.  In our study, we adopt the isothermal halo
density profile \cite{IsoThermal}
\begin{eqnarray}
  \rho_{\rm DM}^{\rm (Galaxy)} (r) 
  =
  \rho_\odot \frac{r_{\rm core}^2 + r_\odot^2}{r_{\rm core}^2+r^2},
  \label{eq:isothermal}
\end{eqnarray}
where $\rho_\odot\simeq 0.43\ {\rm GeV/cm^3}$ is the local halo density,
$r_{\rm core}\simeq 2.8\ {\rm kpc}$ is the core radius, $r_\odot\simeq
8.5\ {\rm kpc}$ is the distance between the Galactic center and the
solar system, and $r$ is the distance from the Galactic center.  In
studying the propagation of the cosmic-ray $e^\pm$, we adopt the shape
of the diffusion zone used in the so-called MED propagation model
\cite{Delahaye:2007fr}; the diffusion zone is approximated by a cylinder
with the half-height of $L=4\ {\rm kpc}$ and the radius of $R=20\ {\rm
kpc}$.  Notice that the solution of the diffusion equation is
insensitive to the diffusion coefficient $K_e(E)$ for high energy
$e^{\pm}$ \cite{Ishiwata:2008qy}, as we will discuss in the following.
In addition, the energy loss rate $b_{\rm loss}$ is given by the sum of
the contributions from the synchrotron-radiation and the IC processes:
$b_{\rm loss}^{\rm (Galaxy)}(E_e,\vec{x})=b_{\rm synch}^{\rm
(Galaxy)}(E_e,\vec{x})+b_{\rm IC}^{\rm (Galaxy)}(E_e,\vec{x})$.  For the
calculation of $b_{\rm synch}^{\rm (Galaxy)}$, we approximate that the
strength of the magnetic flux density $B$ is independent of position in
our Galaxy; then we obtain
\begin{eqnarray}
  b_{\rm synch}^{\rm (Galaxy)}(E_e) = 
   \sigma_{\rm T} \gamma_e^2 B^2,
\end{eqnarray}
with $\gamma_e=E_e/m_e$ and $\sigma_{\rm T}$ being the cross section of
the Thomson scattering.  In our numerical study, we use $B=3\ \mu{\rm
G}$. Furthermore, $b_{\rm IC}^{\rm (Galaxy)}$ is given by
\begin{eqnarray}
  b_{\rm IC}^{\rm (Galaxy)} (E_e, \vec{x}) = c
  \int dE_{\gamma} dE_{\gamma_{\rm BG}} 
  (E_{\gamma}-E_{\gamma_{\rm BG}})
  \frac{d\sigma_{\rm IC}}{dE_\gamma}
  f_{\gamma_{\rm BG}} (E_{\gamma_{\rm BG}}, \vec{x}),
\end{eqnarray}
where $c$ is the speed of light and the differential cross section for
the IC process is expressed as \cite{Blumenthal:1970gc}
\begin{eqnarray}
  \frac{d\sigma_{\rm IC}}{dE_\gamma}
   =
  \frac{3\sigma_{\rm T}}{4\gamma_e^2 E_{\gamma_{\rm BG}}}
  \left[
    2 q \ln q + (1 + 2q) (1 - q) 
    + \frac{(\Gamma_e q)^2 (1 - q)}{2 (1 + \Gamma_e q)}
  \right],
\end{eqnarray}
with $\Gamma_e=4\gamma_eE_{\gamma_{\rm BG}}/m_e$, $q=E_\gamma/\Gamma_e
(E_e - E_\gamma)$, and $f_{\gamma_{\rm BG}}$ is the spectrum of the
background radiation photon.  Kinematically, $1/4\gamma_e^2\leq q\leq 1$
is allowed.  The background photon in our Galaxy has three components:
(i) star light concentrated in the Galaxy, (ii) star light re-scattered
by dust, and (iii) the CMB radiation.  The spectrum of the CMB radiation
is isotropic and well known, while those of the first and second
components depend on the position.  We use the data of interstellar
radiation field provided by the GALPROP collaboration \cite{Galprop},
which is based on \cite{Porter:2005qx}, to calculate $f_{\gamma_{\rm
BG}}$ in our Galaxy.

The typical propagation length of electron per time scale of the
energy loss is estimated to be $O(0.1\ {\rm kpc})$ for $E_e\sim 100\
{\rm GeV}$, and it becomes shorter as the energy increases.  (Here, we
have used the diffusion coefficient suggested in the MED propagation
model, $K(E)=0.0112\ {\rm kpc^2/Myr}\times (E_e/1\ {\rm
  GeV})^{0.70}$.)  Then, the $e^\pm$ spectrum at the position
$\vec{x}$ is well approximated by
\begin{eqnarray}
  f_e^{\rm (Galaxy)} (E_e,\vec{x})
  = \frac{1}{b_{\rm loss}^{\rm (Galaxy)} (E_e,{\vec{x}})}
  \frac{\rho_{\rm DM}^{\rm (Galaxy)} (\vec{x})}{\tau_{\rm DM} m_{\rm DM}}
  \int_{E_e}^{\infty} dE_e^{\prime} \frac{dN_e}{dE_e^{\prime}}.
  \label{f_e(Gal)}
\end{eqnarray}
In our numerical analysis, we adopt the above approximated formula for
the $e^\pm$ spectrum in our Galaxy.  Then, $\gamma$-ray flux from
direction ($b,l$), where $b$ and $l$ are Galactic latitude and
longitude, respectively, is obtained by line-of-sight (l.o.s) integral
of $\gamma$-ray energy density per unit time and unit energy as
\begin{eqnarray}
  \Phi_{\gamma}^{\rm (Galaxy)} (b,l)=
  \frac{1}{4 \pi} \int_{\rm l.o.s} d \vec{l} 
  L_{\rm IC}(E_{\gamma}, \vec{l}),
\end{eqnarray}
where  
\begin{eqnarray}
  L_{\rm IC} (E_\gamma, \vec{x})
  = c \int d E_e d E_{\gamma_{\rm BG}} 
  \frac{d\sigma_{\rm IC}}{dE_\gamma}
  f_{\gamma_{\rm BG}} (E_{\gamma_{\rm BG}}, \vec{x})
    f_e^{\rm (Galaxy)} (E_e, \vec{x}).
\end{eqnarray}

In studying the $\gamma$-ray from cosmological distance, we need to
understand the $e^\pm$ spectrum in the extra-Galactic region.  In such
a region, the $e^\pm$ spectrum is independent of the position, so the
spectrum should obey
\begin{eqnarray}
  \frac{\partial f_e(t, E_e)}{\partial t} =
  H E_e \frac{\partial f_e(t, E_e)}{\partial E_e}
  + \frac{\partial}{\partial E_e}
  \left[ b_{\rm loss}(t, E_e) f_{e}(t, E_e) \right]
  + Q(t, E_e),
  \label{Boltzmann_C}
\end{eqnarray}
where $H$ is the expansion rate of the universe.  Contrary to the case
in our Galaxy, only the IC process with the CMB radiation contributes to
the energy-loss process.  Since the typical energy of the CMB radiation
is so low that $e^\pm$ becomes non-relativistic in the center-of-mass
energy of the IC process.  Then, taking into account the red-shift of
the CMB radiation, the energy loss rate is given by
\begin{eqnarray}
  b_{\rm loss}^{\rm (Cosmo)} (t, E_e) = 
  \frac{4}{3} \sigma_{\rm T} \gamma_e^2 \rho_{\rm CMB}^{\rm (now)} 
  (1+z)^4,
\end{eqnarray}
where $\rho_{\rm CMB}^{({\rm now})}\simeq 0.26\ {\rm eV/cm^3}$ is the
present energy density of the CMB, and $z$ is the red-shift.

The typical time scale of the energy loss due to the IC process is
estimated to be $E_e/b_{\rm loss}^{\rm (Cosmo)}$, and is of the order of
$10^{14}\ {\rm sec}$ for $E_e=100\ {\rm GeV}$ (and becomes shorter as
$E_e$ increases).  Because the energetic $e^\pm$ loses its energy via
the IC process before the energy is red-shifted, we neglect the terms of
$O(H)$ in Eq.\ \eqref{Boltzmann_C} and obtain
\begin{eqnarray}
  f_e^{({\rm Cosmo})} (t, E_e)
  = \frac{1}{b_{\rm loss}^{({\rm Cosmo})}(t, E_e)}
  \frac{\rho_{\rm DM}^{({\rm now})}(1+z)^3}{\tau_{\rm DM} m_{\rm DM}}
  \int_{E_e}^{\infty} dE_e^{\prime} \frac{dN_e}{dE_e^{\prime}},
  \label{f_e(Cosmo)}
\end{eqnarray}
where $\rho_{\rm DM}^{({\rm now})}\simeq 1.2\times 10^{-6}\ {\rm
  GeV}/{\rm cm}^3$ is the present energy density of dark matter.
Then, taking into account red-shift of scattered photon spectrum
and the dilution due to the expansion of the universe, we obtain
the $\gamma$-ray from cosmological distance as
\begin{eqnarray}
  \Phi_{\gamma}^{\rm (Cosmo)} = 
  \frac{c}{4 \pi} \int dt \frac{1}{(1+z)^3} L_{\rm IC}(t,E_{\gamma}),
\end{eqnarray}
where
\begin{eqnarray}
  L_{\rm IC}(t,E_{\gamma})
  = 
  (1+z)c \int d E_e d E_{\gamma_{\rm BG}} 
  \left[
  \frac{d\sigma_{\rm IC}}{dE'_\gamma}
  \right]_{E'_\gamma=(1+z)E_\gamma}
  f_{\gamma_{\rm BG}}^{\rm (CMB)} (t, E_{\gamma_{\rm BG}})
  f_e^{\rm (Cosmo)} (t, E_e),
   \nonumber \\
\end{eqnarray}
with $f_{\gamma_{\rm BG}}^{\rm (CMB)} (t, E_{\gamma_{\rm BG}})$ being
the spectrum of the CMB radiation at the time $t$.  Notice that the
astrophysical uncertainty is small in the extra-Galactic contribution,
as is obvious from the above expression.

\section{Numerical results}

Now, we are at the position to show our numerical results.  First we
will discussthe case where dark matter dominantly decays into
$\mu^+\mu^-$ pair as an example.  In such a case, energetic $e^\pm$ are
produced by the decay of $\mu^\pm$.  Then, the positron fraction can be
in good agreement with the PAMELA result while the total $(e^++e^-)$
flux can be consistent with the Fermi data if $m_{\rm DM}\sim O(1~{\rm
TeV})$ and $\tau_{\rm DM}\sim O(10^{26}\ {\rm sec})$. In order to choose
best-fit value with PAMELA data for $\tau_{\rm DM}$ in a given $m_{\rm
DM}$, we calculate $\chi^2$. In this anlysis, we simulate background
flux of $e^{\pm}$ by the use of GALPROP code \cite{Galprop}.  In the
simulation, we adopt ``conventional'' model which is presented in
\cite{Strong:2004de}.

\begin{figure}
 \centerline{\epsfxsize=0.75\textwidth\epsfbox{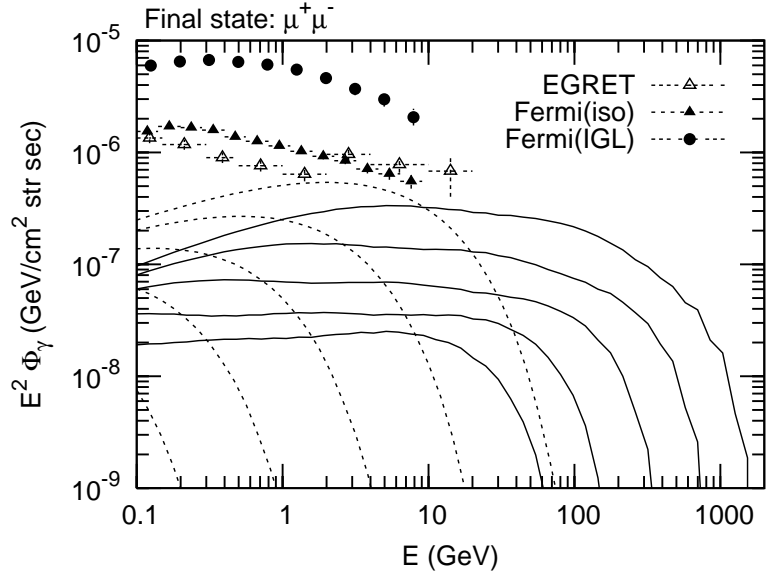}}
   \caption{\small IC $\gamma$-ray flux in the case where dark matter
   decays to $\mu^+\mu^-$ pair. The solid line is the flux from the
   cosmological distance, which is isotropic, while the dashed ones are
   Galactic contributions.  For each line, we take $(m_{\rm
   DM},\tau_{\rm DM})=(250\ {\rm GeV}, 1.0\times 10^{27}\ {\rm sec})$,
   $(500\ {\rm GeV}, 6.2\times 10^{26}\ {\rm sec})$, $(1\ {\rm TeV},
   3.4\times 10^{26}\ {\rm sec})$, $(2\ {\rm TeV}, 1.7\times 10^{26}\
   {\rm sec})$, $(4\ {\rm TeV}, 8.5\times 10^{26}\ {\rm sec})$ from left
   to right. Here, we also plot observation data by EGRET
   \cite{Sreekumar:1997un} and Fermi \cite{Abdo:2009ka,Porter:2009sg}. }
   \label{fig:ic_mumu}
\end{figure}

\begin{figure}
 \centerline{\epsfxsize=0.75\textwidth\epsfbox{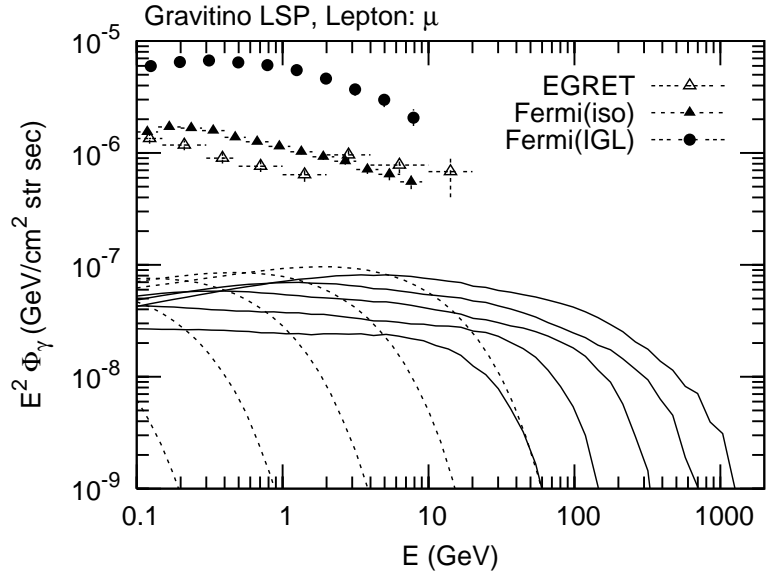}}
   \caption{\small IC $\gamma$-ray flux in the case where gravitino
   dark matter dominantly decays to second generation lepton.  For each
   line, we take $(m_{\rm DM},\tau_{\rm DM})=(250\ {\rm GeV}, 4.0\times
   10^{26}\ {\rm sec})$, $(500\ {\rm GeV}, 2.7\times 10^{26}\ {\rm
   sec})$, $(1\ {\rm TeV}, 2.0\times 10^{26}\ {\rm sec})$, $(2\ {\rm
   TeV}, 1.8\times 10^{26}\ {\rm sec})$, $(4\ {\rm TeV}, 1.6\times
   10^{26}\ {\rm sec})$ from left to right.}  \label{fig:ic_mu}
\end{figure}

\begin{figure}
 \centerline{\epsfxsize=0.75\textwidth\epsfbox{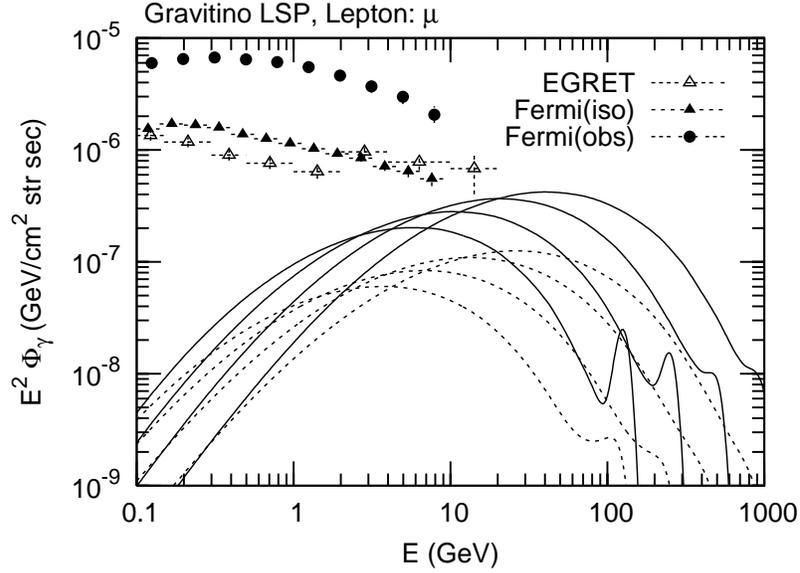}}
   \caption{\small Primarily produced $\gamma$-ray flux in the case
   where gravitino dark matter dominantly decays to second generation
   lepton.  The parameters are taken as the same as Fig.\
   \ref{fig:ic_mu}.}  \label{fig:gm_mu}
\end{figure}

%

In Fig.\ \ref{fig:ic_mumu}, we show the simulated cosmic $\gamma$-ray in
the IC process.  Here, we chose to take $m_{\rm DM}=250~{\rm GeV}$,
$500~{\rm GeV}$, $1~{\rm TeV}$, $2~{\rm TeV}$, and $4~{\rm TeV}$, and
the lifetime is taken as $\tau_{\rm DM} =1.0\times 10^{27}~{\rm sec}$,
$6.2\times 10^{26}~{\rm sec}$, $3.4\times 10^{26}~{\rm sec}$, $1.7\times
10^{26}~{\rm sec}$, and $1.5\times 10^{26}\ {\rm sec}$, respectively.
For Galactic contribution, we plot the averaged flux in
$10^{\circ}<b<20^{\circ}$.  One can see Galactic contribution is much
less than the observed IGL flux.  On the other hand, the flux from
extra-Galactic region makes a large contribution, especially when
$m_{\rm DM}\gtrsim 4\ {\rm TeV}$, and it becomes comparable to the
observed isotropic component of $\gamma$-ray.  Since the isotropic
$\gamma$-ray flux observed by Fermi contains the extra-Galactic
$\gamma$-ray and those from the other sources which are not identified,
the extra-Galactic component should not exceed the observation; thus,
such large dark matter mass may be constrained.

For comparison, we also calculated IC-induced $\gamma$-ray for scenario
in which gravitino $\psi_{\mu}$ becomes dark matter. In the scenario,
gravitino mainly decays to $Wl$, $Z\nu$, and $h\nu$. Then, $e^{\pm}$ is
produced from charged lepton $l$, as well as weak- or higgs bosons; such
$e^{\pm}$ can become a source of the cosmic ray \cite{Ishiwata:2008cu}
to explain PAMELA anomaly \cite{Ishiwata:2008cv,Ishiwata:2009vx}.
Furthermore, total $(e^++e^-)$ flux can be consistent with Fermi
data. Even in this scenario, high energy $\gamma$-ray is inevitably by
$e^{\pm}$ in IC process.  In Fig.\ \ref{fig:ic_mu}, we show IC-induced
$\gamma$-ray in the scenario. Here, we especially consider the case
where gravitino dominantly decays to second-generation lepton, and plot
for $m_{\rm DM}=250~{\rm GeV}$, $500~{\rm GeV}$, $1~{\rm TeV}$, $2~{\rm
TeV}$, and $4~{\rm TeV}$, by choosing best-fit lifetime with PAMELA as
$\tau_{\rm DM} =4.0\times 10^{26}~{\rm sec}$, $2.7\times 10^{26}~{\rm
sec}$, $2.0\times 10^{26}~{\rm sec}$, $1.8\times 10^{26}~{\rm sec}$, and
$1.6\times 10^{26}\ {\rm sec}$, respectively.  One can see that the
Galactic and extra-Galactic fluxes from dark matter is much less than
the present IGL- and isotropic-flux observations.  This result is the
consequence of softer $e^{\pm}$ spectrum from the decay of weak- or
higgs bosons, compared to the previous $\mu^+\mu^-$ case.  Besides, in
$\psi_{\mu}$-DM scenario, the direct production of energetic photon by
the dark matter decay is not negligible.  In the scenario, $\gamma$-ray
is produced directly from the decay of $\psi_{\mu}$, or following weak-
and higgs bosons' decay; those photons also become the source of high
energy cosmic $\gamma$-ray.  In Fig.\ \ref{fig:gm_mu}, we plot the
primarily produced $\gamma$-ray.  Here, we consider the same decay
scenario and parameters as in Fig.\ \ref{fig:ic_mu}.  One can see that
the Galactic and extra-Galactic fluxes from dark matter is much less
than the present IGL- and isotropic-flux observations. Thus, with the
observation data which is currently available, the primarily
$\gamma$-ray is not constrained. However, if data for higher energy
region is obtained, the scenario may be tested or constrained.  For the
other dark matter scenario, the spectrum of the cosmic $\gamma$-ray
depends on the properties of the decaying dark matter.  Detailed
discussion for several specific dark matter models will be given
elsewhere \cite{IshMatMor}.

\section{Conclusion and discussion}

In this paper, we have studied the IC-induced $\gamma$-ray, assuming
that the PAMELA anomaly in the $e^+$ fraction is due to the decay of
dark matter.  We have calculated the Galactic and extra-Galactic
contributions separately, and shown that both of them are important.  In
our study, we have considered the case that dark matter dominantly
decays into $\mu^+\mu^-$ pair, as well as the case of gravitino dark
matter.  However, as far as the PAMELA anomaly is explained in the
decaying dark matter scenario, the IC-induced $\gamma$-ray flux is
expected to be of the same order irrespective of dominant decay process.
This is because, in order to explain the PAMELA anomaly, energetic
$e^\pm$ should be produced by the decay, which induces the IC process.
In our analysis, we found that large dark matter mass over nearly 4 TeV
might be constrained by Fermi observation in the case the dark matter
dominantly decays to $\mu^+\mu^-$ pair, while gravitino dark matter
scenario is not constrained by the present observation.

It should be also noticed that the production rate of energetic $e^\pm$
is extremely suppressed in the extra-Galactic region in another scenario
of explaining the PAMELA anomaly, like the pulsar scenario.
Furthermore, it is notable that the procedure to calculate this
contribution has less uncertainty, and that the flux is isotropic.
Therefore, if detailed data about the $\gamma$-ray spectrum becomes
available for various directions by the Fermi experiment, it will
provide a significant information about the unstable dark matter
scenario.  In particular, study of the $\gamma$-ray flux from directions
away from the Galactic center is important.  For such directions, the
$\gamma$-rays from the Galactic activities are suppressed.  Thus, the
background flux is expected to be comparable to or even smaller than the
extra-Galactic contribution of the IC-induced $\gamma$-ray flux.  

At present, Fermi experiment provides the isotropic $\gamma$-ray, by
analyzing the flux data off Galactic center region \cite{Abdo:2009ka}.
According to their work, however, the isotropic component may contain
contribution from unknown source in Galaxy, as well as extra-Galactic
$\gamma$-ray, and it is mentioned that more detailed analysis is needed
in the future work. Therefore, if the $\gamma$-ray spectrum for
extra-Galactic component is precisely determined, we may see a signal of
the IC-induced $\gamma$-ray. Furthermore, it can be one of the possible
tool for test of dark matter scenario.  Also we note that the
measurement of the flux from directions close to the Galactic center,
which is sensitive to the Galactic contribution, is also important.


\begin{theacknowledgments}
This work was supported in part by Research Fellowships of the Japan
Society for the Promotion of Science for Young Scientists (K.I.), and
by the Grant-in-Aid for Scientific Research from the Ministry of
Education, Science, Sports, and Culture of Japan, No.\ 21740174 (S.M.)
and No.\ 19540255 (T.M.).
\end{theacknowledgments}



\bibliographystyle{aipproc}   

\bibliography{sample}

\IfFileExists{\jobname.bbl}{}
 {\typeout{}
  \typeout{******************************************}
  \typeout{** Please run "bibtex \jobname" to optain}
  \typeout{** the bibliography and then re-run LaTeX}
  \typeout{** twice to fix the references!}
  \typeout{******************************************}
  \typeout{}
 }

\end{document}